\documentstyle[11pt,epsfig]{article}
[1999/03/29 PSNFSS v.7.2
Times font as default roman
: S Rahtz]

\begin{document}
\renewcommand{\thefootnote}{\fnsymbol{footnote}}
\sloppy
\newcommand{\rp}{\right)}
\newcommand{\lp}{\left(}
\newcommand \be  {\begin{equation}}
\newcommand \bea {\begin{eqnarray}}
\newcommand \ee  {\end{equation}}
\newcommand \eea {\end{eqnarray}}

\title{Functional Renormalization Prediction of Rupture}
\thispagestyle{empty}

\author{S. Gluzman$^1$, J.V. Andersen$^1$ and D. Sornette$^{1,2}$\\
$^1$ Laboratoire de Physique de la Mati\`{e}re Condens\'{e}e\\ CNRS UMR6622 and
Universit\'{e} de Nice-Sophia Antipolis\\ B.P. 71, Parc
Valrose, 06108 Nice Cedex 2, France\\
$^2$ Institute of Geophysics and
Planetary Physics and Department of Earth and Space Science\\ 
University of California, Los Angeles, California 90095\\
e-mails: gluz@idirect.com, vitting@unice.fr and sornette@unice.fr \\}

\maketitle
\begin{abstract}
We develop theoretical formulas for the prediction of the rupture of systems
which are known to exhibit a critical behavior, based solely on the
knowledge of the early time evolution of an observable, such as the acoustic
emission rate as a function of time or of stress. From the parameterization
of such early time evolution in terms of a low-order polynomial, we use the
functional renormalization approach introduced by Yukalov and Gluzman to
transform this polynomial into a function which is asymptotically a power
law. The value of the critical time $t_c$, {\it conditioned} on the
assumption that $t_c$ exists, is thus determined from the knowledge of the
coefficients of the polynomials. We test with success this prediction scheme with respect
to the order of the polynomials and as a function of noise.
\end{abstract}

\pagenumbering{arabic}

\section{Introduction}

To what extent can the material failure of a mechanical system under stress
be forecasted? This question has enormous technological interest for its
economic and human consequences, especially in the automobile, naval,
aeronautics and space industries \cite{rocket}, as well as in the sensitive
chemical and nuclear industries to cite a few among many examples.

If rupture occurs brutally without precursors, prediction is impossible. In
contrast, Mogi noticed that, for experiments on a variety of materials, the
larger the heterogeneity of the material, the stronger and more useful are
the precursors to rupture \cite{Mogi1}. For a long time, the Japanese
research effort for earthquake prediction and risk assessment was based on
this very idea \cite{Mogi2}.

These empirical facts have been put on a firm theoretical basis by using
various simplified mechanical models of failure of heterogeneous materials
which showed that, increasing the disorder changes rupture from first-order
(abrupt) to critical (continuous with power-law properties) \cite{Andersen1}%
. By the term ``disorder', we refer to heterogeneity in material properties
(elastic coefficients and rupture thresholds) as well as inhomogeneous
pre-stresses. In the presence of long-range elasticity, disorder is found to
be always relevant leading to a critical rupture. However, the disorder
controls the width of the critical region \cite{thermalfuse4}. The smaller
it is, the smaller will be the critical region, which may become too small
to play any role in practice.

The potential for predicting rupture thus seems associated with its critical
nature. Let us first review past works which support this concept of
critical rupture. Theoretically, the concept that rupture is critical was
first formulated by Vere-Jones \cite{Vere-Jones} using critical branching
models and All\`egre et al. \cite{Allegre} using the percolation model and
real-space renormalization group (see ref.~\cite{book} for a general presentation).
The Russian school has also extensively developed this concept \cite
{Keilis-Borok,Shnirman,ShnrimanBlanter}. One of us and co-workers have
introduced a statistical two-dimensional model of dynamically evolving
damage \cite{thermalfuse}, which exhibits the critical time-to-failure
dependence of the energy released up to the rupture. Based on a precise
numerical description of the many growing interacting micro-cracks with a
spatio-temporal organization which is a function of the stress-dependent
damage law, one finds that, under a step-function stress loading, the total
rate of damage increases on average as a power-law of the time-to-failure.
In this model, rupture is a ``critical point'' in the statistical physics
sense \cite{book} and occurs as the culmination of the progressive
nucleation, growth and fusion between microcracks, leading to a fractal
network of cracks. This simple model has since then been found to describe
quantitatively the experiments on the electric breakdown of
insulator-conducting composites \cite{Lamai} and the damage by
electromigration of polycrystalline metal films \cite{Bradley}. This led to
the proposal and the empirical test on real engineering composite structures
that failure in fiber composites is a genuine ``critical'' point \cite
{Anifrani}. This critical behavior may correspond to an acceleration of the
rate of energy release or to a deceleration, depending on the nature and
range of the stress transfer mechanism and on the loading procedure. A
generalization is to extend the power-law behavior of the time-to-failure
analysis by including corrections in the form of log-periodic modulations 
\cite{Anifrani}. Log-periodicity is the hallmark of a hierarchy of discrete
characteristic scales in the rupture process. Mathematically, it corresponds
to adding an imaginary part to the exponent z (defined below). Intuitively,
the log-periodic oscillations are oscillations that are periodic in the
logarithm of the time-to-failure and thus corresponds to an accelerating
frequency modulation as the critical time is approached. This acceleration
of alternating ups and downs accounts for the succession of damage and
quiescent phases self-organizing and culminating in the rupture.

Following this work \cite{Anifrani}, this method has been used extensively
by the French Aerospace company A\'erospatiale on pressure tanks made of
kevlar-matrix and carbon-matrix composites used on the European Ariane 4 and
5 rockets. In this application, the method consists in recording acoustic
emissions under constant stress rate. The acoustic emission energy as a
function of stress is fitted by the critical theory mentioned above. One of
the parameters is the time of failure and the fit thus provides a
``prediction'' when the sample is not brought to failure in the first test.
The results indicate that a precision of a few percent in the determination
of the stress at rupture is typically obtained using acoustic emission
recorded about 20\% below the stress for rupture. We now have a better
understanding of the conditions, the mathematical properties and physical
mechanisms at the basis of log-periodic structures \cite{reviewdsi}.

The numerical simulations of Sahimi and Arbabi \cite{Arbabi} have confirmed
that, near the global failure point, the cumulative elastic energy released
during fracturing of heterogeneous solids with long-range elastic
interactions exhibit a critical behavior with observable log-periodic
corrections. Molecular dynamics simulations of the geometry of fracture
patterns in a dilute elastic network give similar results \cite{Ray}: under
a uniform strain which drives the fracture to develop by the growth and
coalescence of the vacancy clusters in the network, there exists a
characteristic time at which a dynamical transition occurs with a power law
divergence of the average cluster size. The cluster growth near the critical
time also exhibits spatial scaling in addition to the temporal scaling,
namely as fracture develops with time, the connectivity length of the
clusters increases and diverges at $t_c$. Recent experiments on the rupture
of fiber-glass composites have also confirmed the critical scenario \cite
{Ciliberto}. Johansen and Sornette \cite{critrupt} have recently re-analyzed
the acoustic emissions recorded during the pressurization of spherical tanks
of kevlar or carbon fibers impregnated in a resin matrix wrapped up around a
thin metallic liner (steel or titanium) fabricated and instrumented by
Aerospatiale-Matra Inc. These experiments were performed as part of a
routine industrial procedure, which tests the quality of the tanks prior to
shipment. It was found that the seven acoustic emission recordings of seven
pressure tanks which was brought to rupture exhibit clear acceleration in
agreement with a power-law ``divergence'' expected from the critical point
theory.

At the same time, it became tempting \cite{seismic} to apply similar
considerations to earthquakes. Indeed, over the years there has been a
growing evidence that a significant proportion of large and great
earthquakes are preceded by a period of accelerating seismic activity of
moderate-sized earthquakes. These moderate earthquakes occur during the
years to decades prior to the occurrence of the large or great events and
over a region much larger than its rupture zone. Sornette and Sammis \cite
{seismic} identified a specific measurable signature of this criticality in
terms of a power-law acceleration of the Benioff strain previously
interpreted as an exponential mechanical-damage rate \cite{sykes1,Bufe}. The
combined observational and simulation evidence now seems to confirm that the
period of increased moment release in moderate earthquakes signals the
establishment of long-wavelength correlations in the regional stress field.
Large or great earthquakes appear to dissipate a sufficient proportion of
the accumulated regional strain to destroy these long wavelength stress
correlations \cite{HuangSOCcrit}. They can thus be considered as different
from smaller earthquakes. According to this model, large earthquakes are not
just scaled-up version of small earthquakes but play a special role as
``critical points'' \cite{Bowman,Jaume}. Recent extensions to the intermediate
scale of rockbursts in deep mines confirm the picture \cite{OuillonKn}.

To summarize these works, there is a rather strong evidence that rupture in
heterogeneous media is critical in the sense of statistical physics 
\cite{book}. To what extent can the critical rupture concept be used to predict
rupture? Voight noticed in an exciting precursory work that many systems
fail by exhibiting a typical relationship relating the second time
derivative $d^2\Omega /dt^2$ of some observable $\Omega $ to a positive
power of $\Omega $ itself \cite{Voight,Voight2}. He then used this
relationship to attempt predictions of failures in various materials and of
volcanic eruptions. Basically, the relationship he postulates is nothing but
a power law time-to-failure evolution of the observable $\Omega $
\be
\Omega (t)= ~(t_c-t)^z~, \label{one} 
\ee
where $t_c$ is the critical time of rupture, $z$ is a critical exponent and $%
A$ a numerical amplitude. By differentiating twice and eliminating time, we
indeed get $d^2\Omega /dt^2=B~\Omega ^{\frac{z+2}z}$, where $%
B=z(z+1)~A^{-2/z}$.

As we said above, the critical rupture concept establishes theoretically
this law (\ref{one}) as the result of the cooperative organization of
precursory damage preparing the global rupture. If an observable such as the
rate of acoustic emissions radiated during loading exhibit an acceleration
close to rupture of the form (\ref{one}) as documented in several
experiments \cite{Anifrani,Ciliberto,critrupt}, it is clear that one can try
to fit the data by (\ref{one}) and get a prediction from the determination
of $t_c$.

In practice, the problem is that the fit of a simple power law (\ref{one})
to a noisy data is rather unstable, so much so that often no fits can be
found \cite{kobe}. The problem comes from the fact that only close to $t_c$
(in some relevant time units) can the power law be clearly distinguished
from other parametric accelerating functional forms, such as an exponential.
As an exponential has no critical time, this leads to an ill-defined rupture
time. The fundamental limitation in using this prediction scheme is thus
that $t_c$ is determined only when data is used up to very close to $t_c$.
This is thus far from the prediction goal to infer $t_c$ from a distance!
One possibility to improve the reliability and range of power law fits far
from $t_c$ has been proposed, based on log-periodic corrections to power
laws \cite{Anifrani,reviewdsi,critrupt,seismic,OuillonKn}.

In the present paper, we propose a different approach to the prediction of $%
t_c$. In the next section, we give the gist of the method and formulate the
problem in precise terms. We also provide a brief summary of the functional
renormalization method. In section 3, we test the method in the situation
where both the polynomial expansion at early times and the value of $t_c$ is
known, to see how the time-dependence is reconstructed by the functional
renormalization. In section 4, we present the genuine prediction scheme
which determines $t_c$ solely from the knowledge of the first few
coefficients of the polynomial fit at early times and the assumption that
the late time dynamics is of the power law form (\ref{one}). In particular,
we test how the precision of the prediction improves by adding more terms in
the polynomial expansion. Section 5 presents numerical tests of our method
compared to direct fits by a power law for noisy data and section 6 concludes.

\section{Formulation of the problem and outline of the method}

The gist of the method is as follows. We assume that we are able to
parameterize the early time evolution of an observable, such as the acoustic
emission rate as a function of time or of stress by fitting the data to a
low-order polynomial $\Omega (t)=a_0+a_1t+a_2t^2+...$. We then use the
sophisticated functional renormalization approach introduced by Yukalov and
Yukalov and Gluzman \cite{G1}-\cite{G12}. We use the version developed in 
\cite{G8} that transforms a polynomial into an analytic effective sum with
an asymptotic power law of the form (\ref{one}) close to some $t_c$ to be
determined. The value of the critical time $t_c$, {\it conditioned} on the
assumption that $t_c$ exists, is thus determined from the knowledge of the
coefficients $a_0,a_1,a_2,...$

In order to test our proposed scheme, we compare the results of our method
to the exact evolution with time of the macroscopic crack length in an
exactly solvable model of rupture with damage \cite{Rabotnov,GluzSor}. In
this self-consistent theory, the growth of a single macroscopic crack is
controlled by cumulative damage dependent on stress history. The damage $D$
accumulates according to the equation $dD/dt\propto \sigma ^m$, where $%
\sigma $ is the local stress proportional to the globally applied stress $%
\sigma _0$ but which takes into account the distortion due to the crack and $%
m$ is the damage exponent. The law describing the growth of the crack, i.e.,
the dynamics of its half-length $a(t)$, is obtained from the following
self-consistent condition: the time it takes from a point at the some
distance $L$ from the crack tip at time $\tau $ for its damage to reach the
rupture threshold $D^{*}$ is exactly equal to the time taken for the crack
to grow from its size at time $\tau $ by an increment $L$ so that its tip
reaches the point exactly when it ruptures. For $m=1$, the full solution is
known \cite{Rabotnov,GluzSor} 
\be
a(t)= {\frac{a_0}{\cos (\pi \sigma _0t/3D^{*})}}~, \label{njaaajakaaak} 
\ee
and is indeed of the form (\ref{one}) close to $t_c=3D^{*}/2\sigma _0$ with $%
z=-1$.

For the convenience of notations, we work in the sequel with dimensionless
variables $f=a/a_0$ and ${\frac{\pi \sigma _0}{3D^{*}}}~t\to t$, so that the
solution (\ref{njaaajakaaak}) becomes 
\be
f_{\rm exact}(t)= {\frac 1{\cos t}}~. \label{njjakaaak} 
\ee

Let us assume that we have access only to the small time dynamics, captured
mathematically by the first terms in the expansion of the solution (\ref
{njjakaaak}): 
\be
f_a(t) = 1 + a_1 t^2 + a_2 t^4 + a_3 t^6+...~~~~~~~~{\rm with}%
~a_1=1/2,~~a_2=5/24~~{\rm and}~~a_3={\frac{61}{720}}~. \label{cfnncx} 
\ee
Since the cosine function is even in $t$, only even power of $t$ are
present. Knowledge of only the first few terms in (\ref{cfnncx}) is the
relevant situation for instance in an experiment in which the early acoustic
emissions are recorded and one would like to infer the subsequent evolution.

To make this paper self-consistent, we first outline the method we use,
which is a direct adaptation of ref.~\cite{G8}. The complete mathematical
foundation can be found in earlier publications \cite{G1}-\cite{G12}. Assume
that we are interested in a function $\phi (x)$ of a real variable $x$. Let
perturbation theory give for this function the perturbative approximations $%
p_k(x)$ with $k=0,1,2,...$, enumerating the approximation order. Define the
algebraic transform $F_k(x,s)=x^sp_k(x)$. This transform changes the powers
of the series $p_k(x)$, thus changing its convergence properties. As a
result, the approximation order effectively increases from $k$ to $k+s$. The
inverse transform is $p_k(x)=x^{-s}F_k(x,s)$. Define the expansion function $%
x=x(f,s)$ by the equation $F_0(x,s)=f$, where $F_0$ is the first available
approximation and $f$ is a new variable. Substituting $x(f,s)$ back to $F_k$%
, we get $y_k(f,s)=F_k(x(f,s),s)$. The transformation inverse to the latter
reads $F_k(x,s)=y_k(F_0(x,s),s)$.

Consider the family $\{y_k\}$ as a dynamical system in discrete time $k$,
the order of the approximations. The trajectory $\{y_k(f,s)\}$ of this
dynamical system is, by construction, bijective to the approximation
sequence $h\{F_k(x,s)\}$. This system can thus be called the approximation
cascade. The next step is to embed the discrete sequence $\{y_k(f,s)\}$ into
a continuous sequence $\{y(\tau ,f,s)\}$ with $\tau \in [0,+\infty ]$. Thus,
the family $\{y(\tau ,f,s):\tau \in [0,+\infty ]\}$ composes a dynamical
system with continuous time, whose trajectory passes through all points of
the approximation cascade trajectory. Such a system can thus be called the
approximation flow. The evolution equation for a flow can be presented in
the functional form $y(\tau +\tau ^{\prime },f,s)=y(\tau ,y(\tau ^{\prime
},f,s),s)$. We call this equation the self-similarity relation, which is the
central concept of our approach. In this framework, the motion occurs in the
space of approximations, where self-similarity is a necessary condition for
convergence as a function of ``time'' defined as the approximation number.
The evolution equation for the approximation flow can be rewritten in the
differential form and then integrated over time between $k$ and some $k^{*}$%
. The point $k^{*}$ is to be chosen to provide the best approximation $%
F_{k+1}^{*}(x,s)=y(k^{*},F_0(x,s),s)$ for the minimal time $k^{*}-k$. The
cascade velocity $v_k(y,s)$ in the vicinity of the time $k$ may be presented
by the Euler discretization of the flow velocity giving $%
v_k(f,s)=V_k(x(f,s),s)$, with $V_k(x(f,s),s)=F_{k+1}(x,s)-F_k(x,s)$. The
integral form of the evolution equation is \be
\int_{F_k}^{F_{k+1}^{*}} {\frac{df}{v_k(f,s)}} = k^{*}-k~, \label{nvbjd} 
\ee
where $F_k=F_k(f,s)$ and $F_{k+1}^{*}=F_{k+1}^{*}(x,s)$. The approximation $%
F_{k+1}^{*}$ must be reached during the minimal time. When no additional
constraints are imposed, then the minimal time corresponds, evidently, to
one step: $k^{*}=k+1$. Finding $F_k^{*}(x,s)$ from (\ref{nvbjd}) and using
the inverse transform leads to the self-similar approximation $%
p_k^{*}(x,s)=x^{-s}F_k^{*}(x,s)$.

Now, by means of the substitution $s\to s_k$, we have to introduce control
functions s$_k$ which can govern the convergence of the sequence $%
\{p_k^{*}(x,s_k)\}$. Following the standard procedure \cite{G1}-\cite{G12}
and similar to the steps described above, we may construct an approximation
cascade with its corresponding cascade velocity. Convergence of an
approximation sequence is, in the language of dynamical theory, the same as
the existence of an attracting fixed point for the corresponding
approximation cascade. If the cascade trajectory tends to a fixed point,
this means that the flow velocity goes to zero as ``time'' $k$ goes to
infinity. In practice, we cannot, of course, reach the limit $k\to \infty $,
and have to stop at a finite (approximation order) $k$. Then, the condition
to be as close to a fixed point as possible is the minimum of the velocity.

After the control functions are found from the minimal-velocity condition,
we substitute them into $p_k^{*}$ and obtain the final expression $%
f_k^{*}=p_k^{*}(x,s_k(x))$ for the self-similar approximation of the sought
function. The practical way of using the minimal-velocity condition is to
express it as a minimal difference condition. To check whether the obtained
sequence $\{f_k^{*}(x)\}$ converges, we have to analyze whether the
corresponding mapping is contracting. The mapping related to the sequence $%
\{f_k^{*}\}$ is constructed in the standard way \cite{G1}-\cite{G12} and the
contraction, or stability, is analyzed by calculating the mapping
multipliers. Ref.~\cite{G8} has shown how this method of algebraic
self-similar renormalization works to obtain accurate estimations of the
critical behavior of a large variety of physical systems, starting from
virial-type or perturbation expansions containing only second-order terms
and derived for a region far from the critical point.

In the sequel, we translate this formalism to the case where $x$ is now a
real time, $k$ is the order of the polynomial fit to the early time of the
signal which plays also the role of ``time'' for the dynamical flow in the
functional space. The goal is to describe as accurately as possible the
finite-time singularity, which is equivalent to a critical point in the time
domain. Before investigating the predictive power of our approach, we first
investigate the possibility of reconstructing as faithfully as possible the
time evolution based on the knowledge of the singularity.

\section{Reconstruction of the crack dynamics from the knowledge of the
small time dynamics and the position of a finite-time singularity}

Here, we assume in addition that we know the position of the singularity at
$t={\frac \pi 2}$ and its exponent, i.e., that 
\be
f(t) = {1 \over \pi /2 - t}~~~~~~{\rm for}~~t \to \pi /2~. \label{nbvvvx} 
\ee
Note that the amplitude $A$ of the pole is exactly equal to $1$. We apply
the Yukalov-Gluzman technique in its version related to crossover phenomena 
\cite{G9}-\cite{G12} , to obtain the best approximation of $f(t)$ for
arbitrary times based only on the knowledge (\ref{cfnncx}) at small times
and (\ref{nbvvvx}) at times close to rupture. 

\subsection{Using only $f_a(t) = 1 + a_1 t^2$ and (\ref{nbvvvx})}

From the expansion $f_a(t)=1+a_1t^2$, the Yukalov-Gluzman method allows us
to build the approximant 
\be
f_1^{*}(t) = \left[ \left( \exp \left( a_1t^2\right) \right) ^{-{\frac
1\beta }}+bt^4\right] ^{-\beta }~,
\ee
with two unknown parameters b and $\beta$, to be determined solely by
demanding the existence of a crossover. The exponent $\beta $ is determined from the
condition that $f_a(t)=1+a_1t^2$ must cross-over to (\ref{nbvvvx}). This
gives $\beta =1$ for the simple pole (\ref{nbvvvx}). The coefficient $b$ is
obtained from the condition of a pole at the known critical point $t_c=\pi
/2 $, which reads $1/f_1(t_c)=0$. Solving for $b$, we finally get the
first-order approximant 
\be
f_1^{*}(t) = \left[ \exp \left( -a_1t^2\right) -{\frac{t^4}{t_c^4}}\exp
\left( -a_1t_c^2\right) \right] ^{-1}~.  \label{eq8}
\ee

\begin{figure}
\epsfig{file=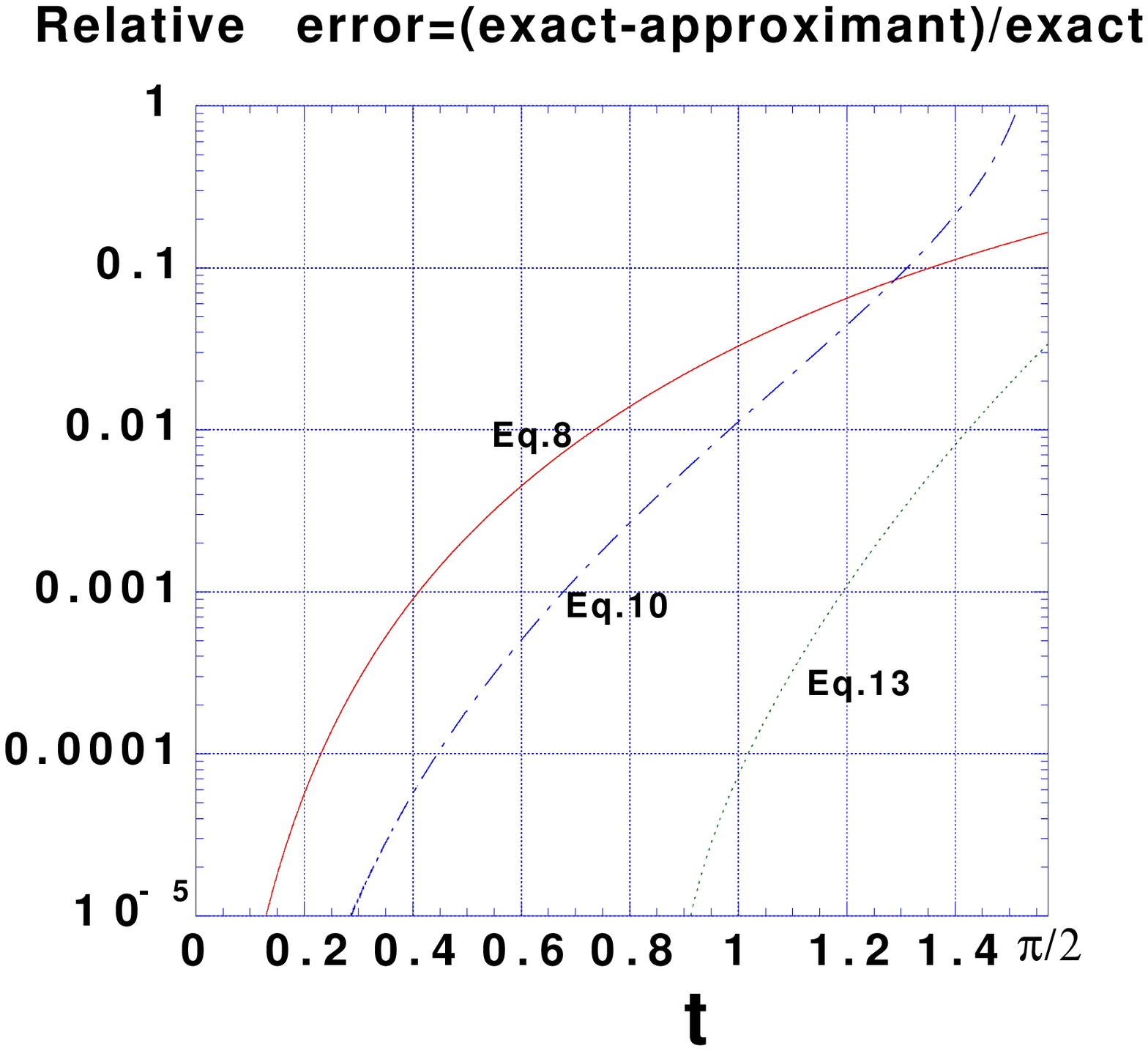,width=16cm}
\caption{\label{fig1}  Relative error $[f_{\rm exact}(t)-f_i^*(t)]/f_{\rm exact}(t)$,
for the three approximations formulas (\ref{eq8}) for $f_1^*(t)$,
(\ref{fjaakkaaa}) for  $f_2^*(t)$, and (\ref{fjaakkaaaa}) for $f_3^*(t)$, where
$f_{\rm exact}(t)$ is given by (\ref{njjakaaak}). 
 }
\end{figure}

This approximant gives an amplitude $A_1=0.834$ for the simple pole, only
17\% off the exact value $A=1$. Figure \ref{fig1} plots in logarithmic scale the
relative errors between the approximant $f_1$ and the exact expression (\ref{njjakaaak})
as a function of time $t$.

\subsection{Using $f_a(t) = 1 + a_1 t^2 + a_2 t^4$ and (\ref{nbvvvx})}

A similar procedure as in the previous case gives the second-order
approximant 
\be
f_2^{*}(t) = \left[ \exp \left( -a_1t^2\exp \left( {\frac{a_2}{a_1}}%
t^2\right) \right) -{\frac{t^6}{t_c^6}}\exp \left( -a_1t_c^2\exp \left( {\ 
\frac{a_2}{a_1}}t_c^2\right) \right) \right] ^{-1}~. \label{fjakka} 
\ee

This second-order approximant can be improved by introducing an additional
control parameter $\tau~$(optimal effective time $k^{*}-k$) such that we can
enforce the condition that the coefficient $a_2$ is preserved in the
renormalization procedure. In other words, the expansion of $f_2(t)$ is now
imposed to have the same coefficient of its power $t^4$. The expression (\ref
{fjakka}) is thus modified into 
\be
f_2^{*}(t) = \left[ \exp \left( -a_1t^2\exp \left( {\frac{a_2}{a_1}~}%
\tau~t^2\right) \right) -{\frac{t^6}{t_c^6}}\exp \left( -a_1t_c^2\exp \left( {\ 
\frac{a_2}{a_1}~}\tau~t_c^2\right) \right) \right]  ^{-1}~. \label{fjaakkaaa} 
\ee

The amplitude of the simple pole predicted by this approximant is found
equal to $0.898$, only 10\% off from the exact value $1$. The condition that
the coefficient of the power $t^4$ in the expansion of (\ref{fjaakkaaa}) is
equal to $a_2=5/24$ given by (\ref{cfnncx}) leads to the improved second
approximant of the form (\ref{fjaakkaaa}) with 
\be
\tau=1-{\frac{a_1^2}{2a_2}}~. 
\ee

Figure \ref{fig1} plots in logarithmic scale the
relative errors between the approximant $f_2$ and the exact expression (\ref{njjakaaak})
as a function of time $t$.

\subsection{Using $f_a(t) = 1 + a_1 t^2 + a_2 t^4 + a_3 t^6$ and (\ref{nbvvvx})}

An extension one step further in the Yukalov-Gluzman procedure gives 
\be
f_3^{*}(t) = \left[ \left( \exp \left( a_1t^2\exp \left( {\frac{a_2}{a_1}~}%
t^2\tau_1\exp \left( {\frac{a_3}{a_2}~}t^2\tau_2\right) \right) \right) \right) ^{-%
{\frac 1\beta }}~+~a_8t^8\right] ^{-\beta }~, 
\ee
where the two control parameters $\tau_1$ and $\tau_2$ are to be determined from
the condition that $a_2$ and $a_3$ are conserved by the renormalization. 
We find
\be
\tau_1 = 1 - {a_1^2 \over 2 a_2}
\ee
and
\be
\tau_2 = -{1 \over a_3 \tau_1} \left( {a_2^2 \tau_1^2 \over 2 a_1} + a_1 a_2 \tau_1 + 
{a_1^3 \over 6} - a_3 \right)~.
\ee
The
coefficient $a_8$ is found from the condition on the critical point:
\be
a_8 = -{1 \over t_c^8} \exp\left(-{a_1 \over \beta} t_c^2 \exp\left({a_2\over a_1}t_c^2 \tau_1
\exp\left({a_3 \over a_2} t_c^2 \tau_2\right)\right)\right)
\ee
The
solution then reads 
\be
f_3^{*}(t) = \left[ \exp \left( -{\frac 12}t^2\exp \left( {\frac 16}t^2\exp
\left( {\frac{11}{60}}t^2\right) \right) \right) -256\left({t \over t_c}\right)^8
\exp \left( -{\frac{\pi ^2}8}\exp \left( {\frac{\pi ^2}{24}}
\exp \left( {\frac{11\pi ^2}{240}}
\right) \right) \right) \right] ^{-1}~. \label{fjaakkaaaa} 
\ee
This approximant gives an amplitude $A_3=0.9661$ for the simple pole, only
3.4\% off the exact value $A=1$. The quality of the reconstruction of the
full function (\ref{njjakaaak}) can also be checked by comparing the exact
value of the next term $(277/8064)~t^8=0.03435~t^8$ in the expansion (\ref
{cfnncx}) of the function (\ref{njjakaaak}). We find $0.034497~t^8$,
corresponding to an error of $0.4\%$.

Figure 1 plots in logarithmic scale the relative errors between the
approximants $f_1^{*}$, $f_2^{*}$ and $f_3^{*}$ and the exact expression (\ref{njjakaaak}).

\section{Prediction of the critical time from the knowledge of the small
time dynamics}

We now assume the knowledge of the expansion (\ref{cfnncx}) up to some
order, representing for instance the experimental recording of an acoustic
emission signal up to some stress level. In addition, we assume {\it only}
the existence of a singularity of the form $1/(t_c-t)^K$ at some value $t_c$%
, using the insight from the theory of critical rupture, but do not know a
priori neither the position of $t_c$, nor the value of the exponent $K$. In
other words, we assume that we know that the system is bound to break but we
do not know when and how. Our goal is to attempt to determine the critical
time and the functional form of the signal on the approach to the critical
rupture time from the recording of the early signals.

\subsection{Standard approach from Yukalov and Gluzman \cite{G8}}

Consider an expansion of some function $\phi $~in powers of some variable $%
u~ $ given by 
\be
p_k(u)=\sum_{k=0}^kb_k~u^k,\quad ~~~~{\rm with}~~ b_0=1~. 
\ee
The method of algebraic self-similar renormalization \cite{G6}-\cite{G8}
gives the following general recurrence formula for the approximant of order $%
k$ as a function of the expansion $p_{k-1}(u)$ up to order $k-1$: 
\be
\phi_k^{*}(u)=p_{k-1}(u)\left[ 1-\frac{k~b_k}s~u^k~p_{k-1}^{k/s}(u)%
\right] ^{-s/k}\equiv \left[ p_{k-1}^{-k/s}(u)-\frac{k~b_k}s~u^k\right]
^{-s/k}~, \label{nvbnkx} 
\ee
where, generally speaking, $s=s_k(u),$ depends on the approximation number
and the variable $u$.

First, let us estimate the position of the critical point. Using only $%
f_1(t)=1+a_1t^2$, we can write it as the inverse of a function that is
requested to vanish in order to obtain the singularity. Expanding in powers
of $t$ up to first order in $t$, we get the estimation $t_{c1}=\sqrt{2}%
=1.414 $, which should be compared with the exact value $\pi /2=1.571$.
Including the next order from the expansion of $\cos (t)^{-1}$ leads to the
improved estimate $t_{c2}=1.59$.

Let us now obtain the expansion which will be used as a raw material for
renormalization. Including the next order in $f_2(t)=f_1(t)+a_2t^4$,
inverting it, and expanding in powers of $t$ up to t$^4-$terms, we obtain an
expansion $p_2(t)$ for $1/f_2:$ 
\be
p_2(t)=\sum_{k=0}^2b_kt^{2k} 
\ee
$$
b_0=1,\quad b_1=-1/2,\quad b_2=1/24, 
$$
Note that, in the initial series, all coefficients in the expansion of $%
cos^{-1}$ are positive, giving the worst possible case for resummation. In
contrast, the coefficients of the inverted series have alternating signs,
which may be better for resummation \cite{G8}. Hereafter, we apply the
resummation procedure to the function $F\equiv 1/f$ inverse of $f$.
Correspondingly, $f^{*}\equiv \left( 1/F\right)^{*}.$

In order to determine the exponent $K$, we follow Yukalov and Gluzman \cite
{G8} and construct the two approximants available from the knowledge of the
two coefficients $a_1$ and $a_2$. They can be readily obtained from the
general formula (\ref{nvbnkx}), with $u=t^2$. The first order approximant is 
\be
F_1^{*}(t)=\left( 1-{\frac{b_1}{s_1}}t^2\right) ^{-s_1} ~. \label{vnsnks} 
\ee
Representing $p_2(t)\ $as $p_2(t)=1+b_1t^2(1+\frac{b_2}{b_1}~t^2)$, and
applying the general formula to the expression in brackets, we obtain the
second order approximant 
\be
F_2^{*}(t)=1+b_1t^2\left( 1-{\frac{b_2}{b_1s_2}}t^2\right) ^{-s_2} ~. 
\label{bvbnnc} 
\ee
Assume further that 
\be
s_{1} = s_2 = s~,  \label{glanvals}
\ee
where $s$ is a single control parameter,
i.e. limit of the total control function in the critical point, which will
play the role of the critical index $K$. As it was explained in \cite{G8},
such an assumption is well justified in the vicinity of a stable fixed point.

We impose the condition of the existence of a critical point, which delivers
two equations for $t_c$ and $s$: 
\be
F_1^{*}(t_c,s)=0~~~~~~{\rm and}~~~~~~F_2^{*}(t_c,s)=0~.  \label{gaknaa}
\ee
The condition of maximum stability of the renormalization amounts to
imposing that the difference $F_2^{*}-F_1^{*}$ be a minimum with respect to
the set of parameters. The minimization of the difference is automatically 
satisfied when (\ref{gaknaa}) holds, since the difference reaches it smallest 
possible value, namely zero.

The vanishing of $F_1^{*}$ given by (\ref{vnsnks})
gives $t_c^2=s/a_1$. The second condition $F_2^{*}=0$ with (\ref{bvbnnc})
yields the estimation $s=1.258~$ for the critical index, only $26\%$ off the
true value equal to $1$. The critical time is given numerically by $%
t_c=1.586 $, very close to the exact value $\pi /2=1.5708$. Note that, as is
often found in critical phenomena, an error of less than $1\%$ in the
location of the critical point is associated with a much larger error of
about $26\%$ on the exponent. In the scheme presented above, it was possible
to find $s$ and $t_c$ from to separate equations.

Such convenience does not hold for another approximation scheme, presented
below, which has however other advantages such as simplicity. In order to
separate the variables, we need an initial guess either for $s$ or for $t_c.$
Such an initial guess is provided naturally by the analog of a mean-field
approximation.

\subsection{Alternative approach: expansion around a ``mean-field''
approximation}

\subsubsection{Expansion to the same order as above}

An alternative and more transparent approach is first to minimize the
distance between approximants and then to verify that (\ref{gaknaa}) holds.
This is the reverse order to the previous scheme that solves (\ref{gaknaa})
which then automatically ensures that the distance between the two successive
approximant is minimized.

In practice, this is implemented as follows. From (\ref{vnsnks}), 
we see that the critical point is located at $t_c^2=s/b_1$, i.e., 
$s=b_1t_c^2$. Using only
\be
p_1(t)=1+b_1t^2~, 
\ee
we estimate $t_c^2=-1/b_1$, which then yields $s=-1$. Note that this value $%
-1$ for the exponent always holds for any value of $b_1$. This exponent thus
plays a role analog to a mean-field approximation in statistical physics.
The fact that the exponent is in the present case equal to the exact value
is a mere coincidence.

In the next order, 
\be
p_2(t)=1+b_1t^2+b_2t^4 ~. 
\ee
The two approximants can be derived directly from the general formulas: 
\be
F_1^{*}(t)=\left( 1-{\frac{b_1}{s_1}}t^2\right) ^{-s_1} ~, 
\ee
\be
F_2^{*}(t)=\left[ \left( 1+b_1t^2\right) ^{-2/s_2}-{\frac{2b_2}{s_2}}%
t^4\right] ^{-s_2/2} ~. 
\ee
We assume in addition that 
\be
s_1=\frac{s_2}{2}=s~,  \label{gbaklaa}
\ee
thus eliminating a trivial dependence of the control parameter $s$ on the
approximation number $k$. Note that (\ref{gbaklaa}) is different from 
(\ref{glanvals}) because we use a different sequence of 
approximations $f_1, f_2,...$. In particular, the condition (\ref{gbaklaa})
ensures that $F_1^{*}(t)$ and $F_2^{*}(t)$ have the same exponent/control
parameter.

In constrast with the previous method of section 4.1, we first minimize the 
difference $F_2^{*}(t,s)-F_1^{*}(t,s)$, and then verify that (\ref{gaknaa}) holds.
The difference calculated at the ``mean-field''
threshold $t_0^2=-1/b_1$ gives 
\be
D_1(s)=\left[ \frac{1+s}s\right] ^{-s}-\left( \frac{-b_2}{b_1^2~s}\right)
^{-s} ~, 
\ee
which is exactly zero at 
\be
s=-1-\frac{b_2}{b_1^2~}=-1.167~. 
\ee
The position of the critical point $t_c$ can be re-calculated from the
condition $F_1^{*}(t_c,s)=0$, which has a non-trivial solution at 
\be
t_c=\sqrt{\frac s{b_1}}=\sqrt{\frac{-1}{b_1}}~\sqrt{1+\frac{b_2}{b_1^2~}}
=1.528~. 
\ee
Expanding in powers of $\frac{b_2}{b_1^2~},$ we estimate $t_c\approx \sqrt{
\frac{-1}{b_1}}\left( 1+\frac{b_2}{2~b_1^2~}\right) =1.532.$

Thus, the renormalization scheme used to calculate $t_c$ and $K$ corresponds
to an expansion around the mean field value $t_c^2=1/b_1$ and $K=1$ in inverse powers
of the dimensionless ``Froude'' number $\frac{b_1^2}{b_2}$ (see \cite
{Techanal} for a definition and use of the Froude number in this context of
functional renormalization). In summary, we get the predictions $t_c=1.528$
closer to the exact value $\pi /2=1.5708,$ and the critical exponent $K=1.167$.

\paragraph{Higher-order expansion}

Including the next order in $f_3(t)=f_2(t)+a_3t^6$, inverting it, and
expanding in powers of $t$ up to $t^6-$terms, we obtain an expansion $p_3(t)$
for $1/f_3$: 
\be
p_3(t)=\sum_{k=0}^3b_kt^{2k} ~, 
\ee
with 
\be
b_1=-1/2,\quad b_2=1/24,\quad b_3=-1/720~. 
\ee
The two higher-order approximants can be written as follows: 
\be
F_2^{*}(t,s_2)=\left[ \left( 1+b_1t^2\right) ^{-2/s_2}-{\frac{2~b_2}{s_2}}%
t^4\right]  ^{-s_2/2} ~, 
\ee
\be
F_3^{*}(t,s_3)=\left[ \left( 1+b_1t^2+b_2t^4\right) ^{-3/s_3}-{\frac{3~b_3 
}{s_3}}t^6\right] ^{-s_3/3} ~. 
\ee
Assume that 
\be
\frac{s_2}2=\frac{s_3}3=s~. 
\ee
The difference $F_3^{*}(t,s)-F_2^{*}(t,s)$ calculated at the ``mean-field''
critical point $t_0^2=-1/b_1$ gives 
\be
D_2(s)=\left[ \left( \frac{b_2}{b_1^2~}\right) ^{-1/s}-\frac{b_3}s\left(
-\frac 1{b_1}\right) ^3\right]  ^{-s}-\left( \frac{-b_2}{b_1^2~s}\right) 
^{-s} ~, 
\ee
which has a zero at $s=-1.024$. The position of the critical point $t_c$ can
be re-calculated from the condition $1/f_3^{*}(t_c,s)=0$, which has a
non-trivial solution at $t_c=1.5722$ determined from the equation 
\be
\left( 1+b_1t_c^2+b_2t_c^4\right) ^{1/s}-{\frac{~b_3}s}t_c^6=0\quad
~~~~~~~(s=-1.024)~, 
\ee
which wins over the ``trivial solution'' of $1+b_1t^2+b_2t^4=0$ at 1.592.
Note that at order $2$, from the condition $F_2^{*}(t_c,s)=0$, we could only
find the trivial solution $t_c^2=1/b_1$. It is only at higher order, starting
with the order $3$ discussed here, that we get corrections to the
``mean-field'' approximation.

To test the validity of the expansion and the strength of the corrections to
the mean-field approximation, let us represent $s$ as $s=-1+X$, substitute
it into equation $D_2(s)=0$, and expand in powers of $X$, thus assuming that 
$X$ is small compared to $-1$. Keeping only terms linear in $X$, $X$ can be
expressed as follows: 
\be
X=-\frac{b_3}{b_1b_2}\frac 1{1-\ln \left( \frac{b_2}{b_1^2}\right) }~, 
\ee
Such an expansion is justified only when $\frac{b_3}{b_1b_2}\ll 1$ and $%
1-\ln \left( \frac{b_2}{b_1^2}\right) $ is significantly different from
zero. In our case, $\frac{b_3}{b_1b_2}=0.067,$ $1-\ln \left( \frac{b_2}{b_1^2%
}\right) =2.792$ and thus $X=-0.024$ is small as expected.

Similarly, let us represent $t_c^2$ as $t_c^2=-1/b_1+C$, substitute it into
the equation $F_3^{*}(t_c,s)=0$, and expand in powers of $C$, thus assuming
that $C$ is small compared to $-1/b_1$. Keeping only the terms linear in $C$%
, we get $C=0.458$, which leads to the value of the critical point $%
t_c=1.568 $ (close to $1.5722$). The expression for $C$ has a simpler
structure at the mean-field point $s=-1$: 
\be
C=-\frac 1{b_1}\left( \frac{b_3}{b_1~b_2}-1\right) \frac 1{2-3\frac{b_3}{%
b_1~b_2}-\frac{b_1^2}{b_2}}~, 
\ee
with an estimate for the critical time given by 
\be
t_c\approx \sqrt{-\frac 1{b_1}}\sqrt{1+\left( \frac{b_3}{b_1b_2}-1\right)
\frac 1{2-3\frac{b_3}{b_1~b_2}-\frac{b_1^2}{b_2}}}=1.563 ~. 
\ee

Thus, the renormalization scheme used to calculate $t_c$ and $K$ corresponds
to an expansion around the mean field value $t_c=1/b_1$ and $K=1,$ in powers
of two dimensionless parameters $\frac{b_3}{b_1~b_2}$ and $\frac{b_2}{b_1^2}$.

In summary, we get the predictions $t_c=1.5722$ very close to the exact
value $\pi /2=1.5708$ and $K=1.024$ only $2\%$ off the exact value $1$.

We conclude that the applicability of the scheme presented in this section
crucially depends on existence of typical small parameters, while the original
scheme of \cite{G8} does not have such a dependence and can be applied even
if critical indices deviate strongly from the mean-field values.

\section{Synthetic tests in the presence of noise}

\begin{figure}
\epsfig{file=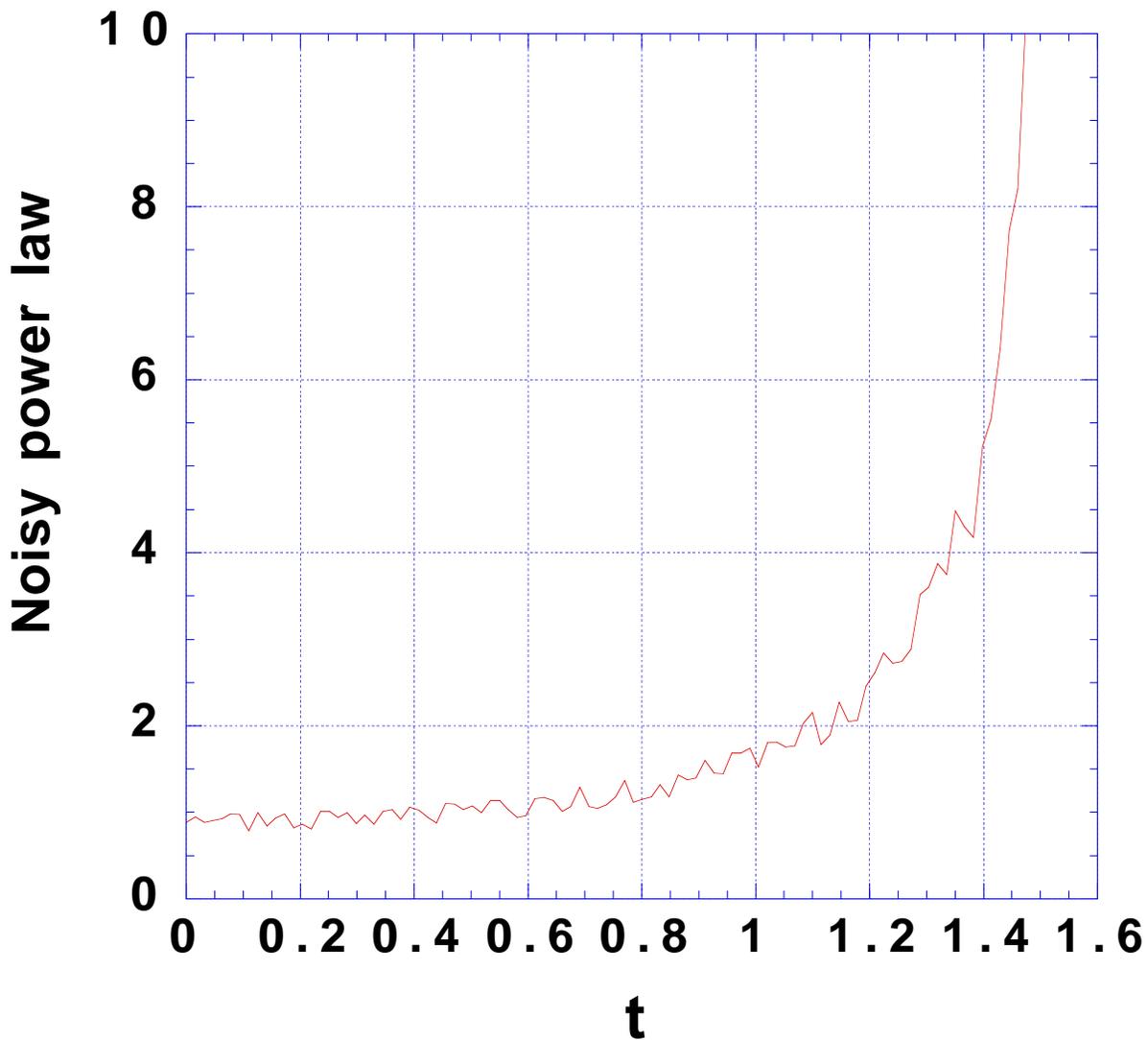,width=16cm}
\caption{\label{fig2} Noisy data obtained from (\ref{njjakaaak}) with
a multiplicative noise of variance  $10^{-2}$. The goal is to use 
this noisy time series up to a maximum value away from $t_c$ to guess what is the critical
value $t_c$ of divergence (theoretically equal to $\pi/2=1.57...$).
 }
\end{figure}

Figure \ref{fig2} shows one realization of a synthetic 
noisy data obtained from (\ref{njjakaaak}) with
a multiplicative noise of variance  $10^{-2}$. This noisy data simulates
an experiment recording a signal as a function of time or of increasing
strain or stress. Our goal is to use 
this noisy time series up to a maximum value away from $t_c$ to guess using
the functional renormalization method what is the critical
value $t_c$ of divergence (theoretically equal to $\pi/2=1.57...$).

\begin{figure}
\epsfig{file=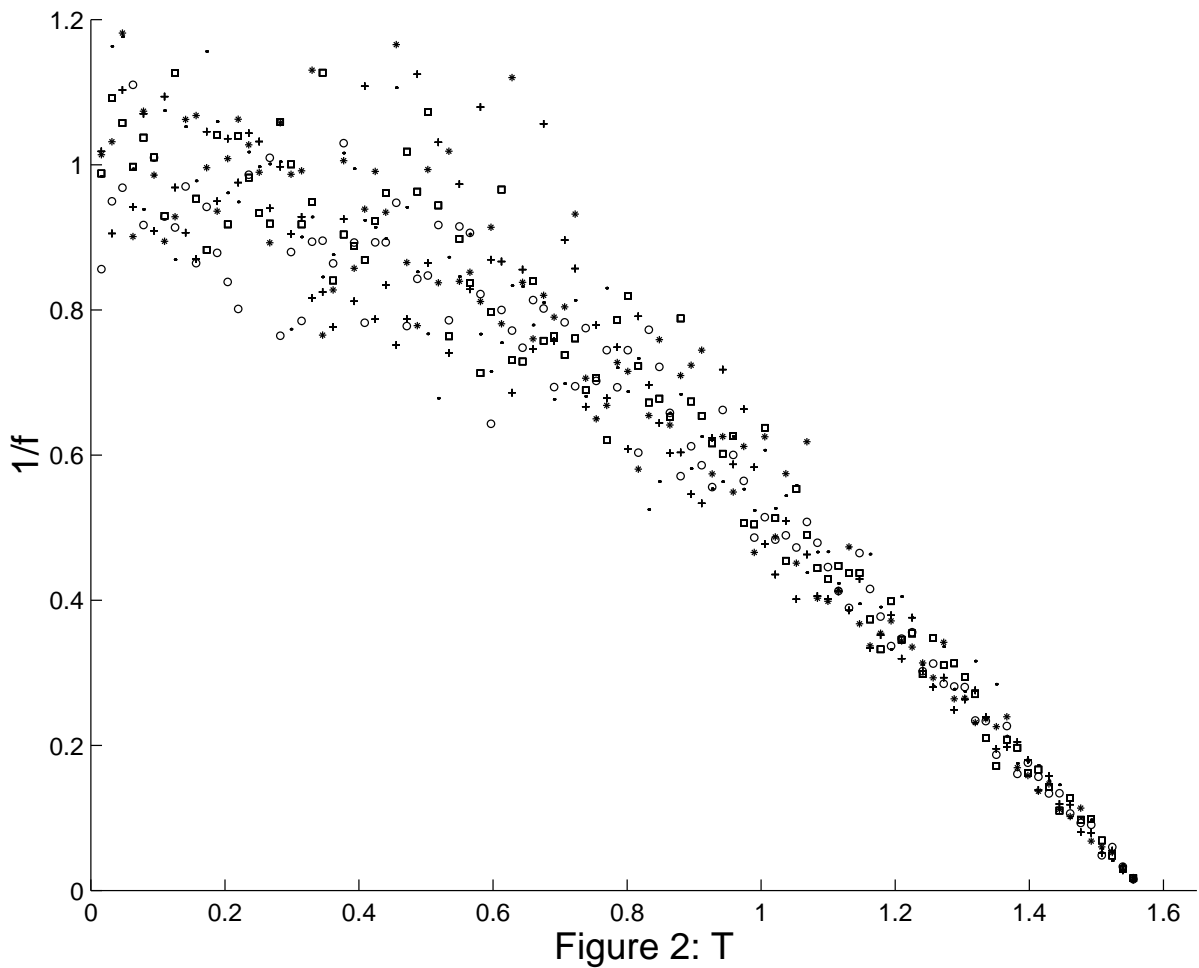,width=16cm}
\caption{\label{fig3} $1$ over the function shown in figure 2 as well
as four other realizations, which are used in
the functional renormalization scheme developed here to predict $t_c$.
The five different symbols shown in figure \ref{fig3}  correspond each to a single 
time series for a specific noise realization. The spread around the 
theoretical $\cos t$ formula gives a sense of the amplitude of the 
multiplicative noise.
 }
\end{figure}

Figure \ref{fig3} shows the inverse of the function in figure \ref{fig2}, 
as well as four other realizations,
which are used in the functional renormalization scheme developed 
in the previous sections to predict $t_c$. The five different
symbols shown in figure \ref{fig3}  correspond each to a single 
time series for a specific noise realization. The spread around the 
theoretical $\cos t$ formula gives a sense of the amplitude of the 
multiplicative noise.

\begin{figure}
\epsfig{file=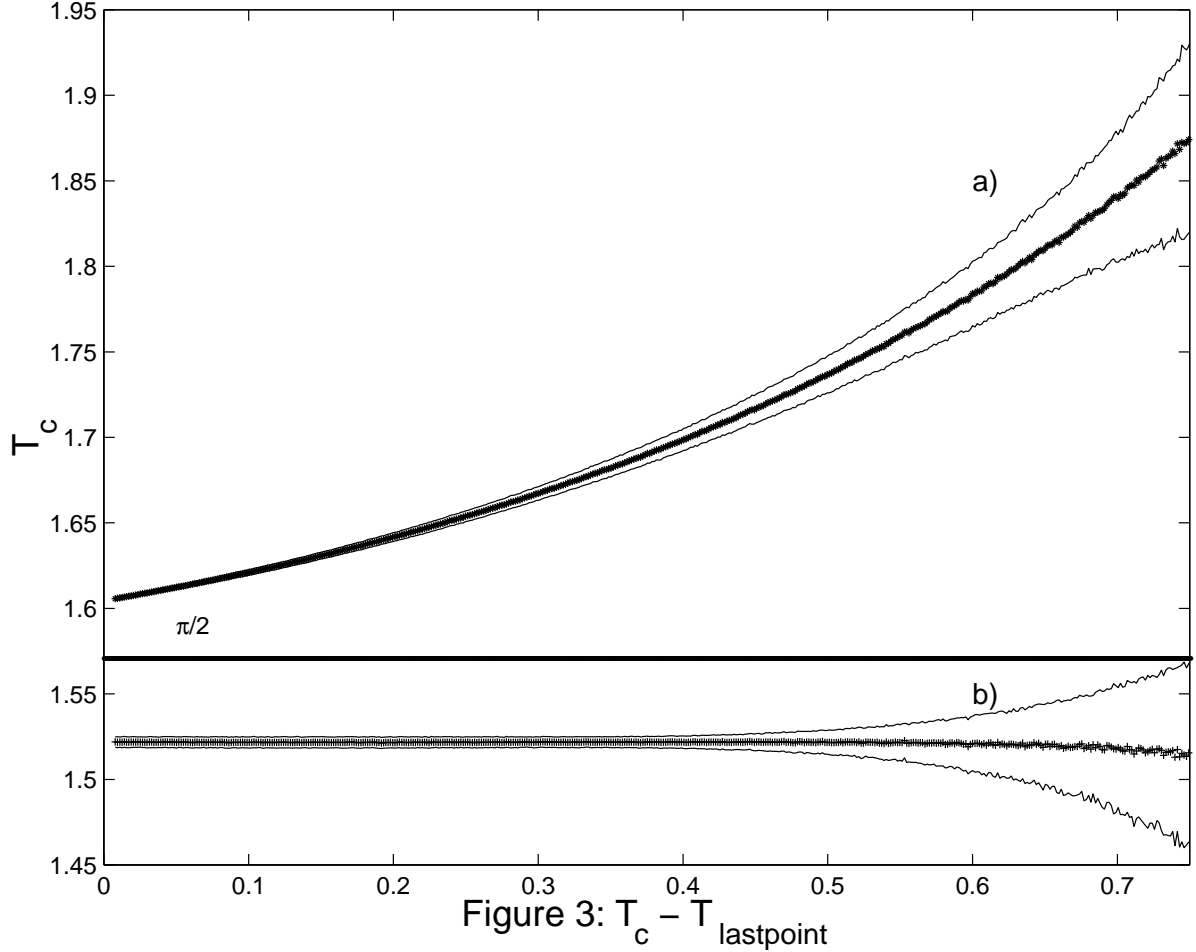,width=16cm}
\caption{\label{fig4}  This figure shows the predicted value for the critical
time $t_c$ as a function of the distance to $t_c$,
 obtained with different schemes, in order to evaluate the
value of the resummation procedures proposed here. In this goal,
1,000 synthetic data sets
were generated using expression (\ref{njjakaaak}) and modifying it
with a multiplicative noise of 
variance $10^{-3}$ corresponding to a standard deviation of $3.3\%$, i.e., with
the equation $f_i(t)= {\frac 1{\cos t}}[1+\eta]$, where $\eta$ is a Gaussian
white noise with variance $10^{-3}$. 
To generate curve a), each data set was fitted by the power law equation 
$A (t_c-t)^{-\beta}$, with $\beta=1$ fixed and $A$ and $t_c$ determined from the fit
in the time interval $[0.5; T_{\rm lastpoint}]$.
The thick line is the average $t_c$ taken over the 1,000 realizations and the two
thin lines are the $\pm$ one standard deviations.
The curve b) is the predicted $t_c$ obtained from our resummation technique
given in section 4.2 which assumes that the exponent 
$\beta$ is close to $1$, using the coefficients $b_1$ and $b_2$ of the fit with the expansion 
$1+b_1t^2+b_2t^4$ to the inverse of each of the synthetic data set $f_i(t)$.
The thick line is the average predicted $t_c$ and the two thin lines are the $\pm$
one standard deviations. The horizontal line at $t_c=\pi/2$ is the exact theoretical
value for the critical time. Up to very close to the critical time, our 
resummation method is clearly superior to the power law fit, even when knowing 
a priori the value of the exponent.}
\end{figure}

Figure \ref{fig4} compares the prediction skill of our procedure described
in section 4.2 to that from a direct fit with a power law. Specifically,
we plot the predicted value for the critical
time $t_c$ as a function of the distance to $t_c$,
 obtained with the two schemes. In this goal, we generated
1,000 synthetic data sets using expression (\ref{njjakaaak}) and modifying it
with a multiplicative noise of 
variance $10^{-3}$ corresponding to a standard deviation of $3.3\%$, i.e., with
the equation $f_i(t)= {\frac 1{\cos t}}[1+\eta]$, where $\eta$ is a Gaussian
white noise with variance $10^{-3}$. 
To generate curve a) in figure \ref{fig4},
each data set was fitted by the power law equation 
$A (t_c-t)^{-\beta}$, with $\beta=1$ fixed and $A$ and $t_c$ determined from the fit
in the time interval $[0.5; T_{\rm lastpoint}]$.
The thick line is the average $t_c$ taken over the 1,000 realizations and the two
thin lines are the $\pm$ one standard deviations.
The curve b) in figure \ref{fig4} is the predicted $t_c$ obtained from our resummation technique
given in section 4.2 which assumes that the exponent 
$\beta$ is close to $1$, using the coefficients $b_1$ and $b_2$ of the fit with the expansion 
$1+b_1t^2+b_2t^4$ to the inverse of each of the synthetic data set $f_i(t)$.
The thick line is the average predicted $t_c$ and the two thin lines are the $\pm$
one standard deviations. The horizontal line at $t_c=\pi/2$ is the exact theoretical
value for the critical time. Up to very close to the critical time, our 
resummation method is clearly superior to the power law fit, even when knowing 
a priori the value of the exponent.

\begin{figure}
\epsfig{file=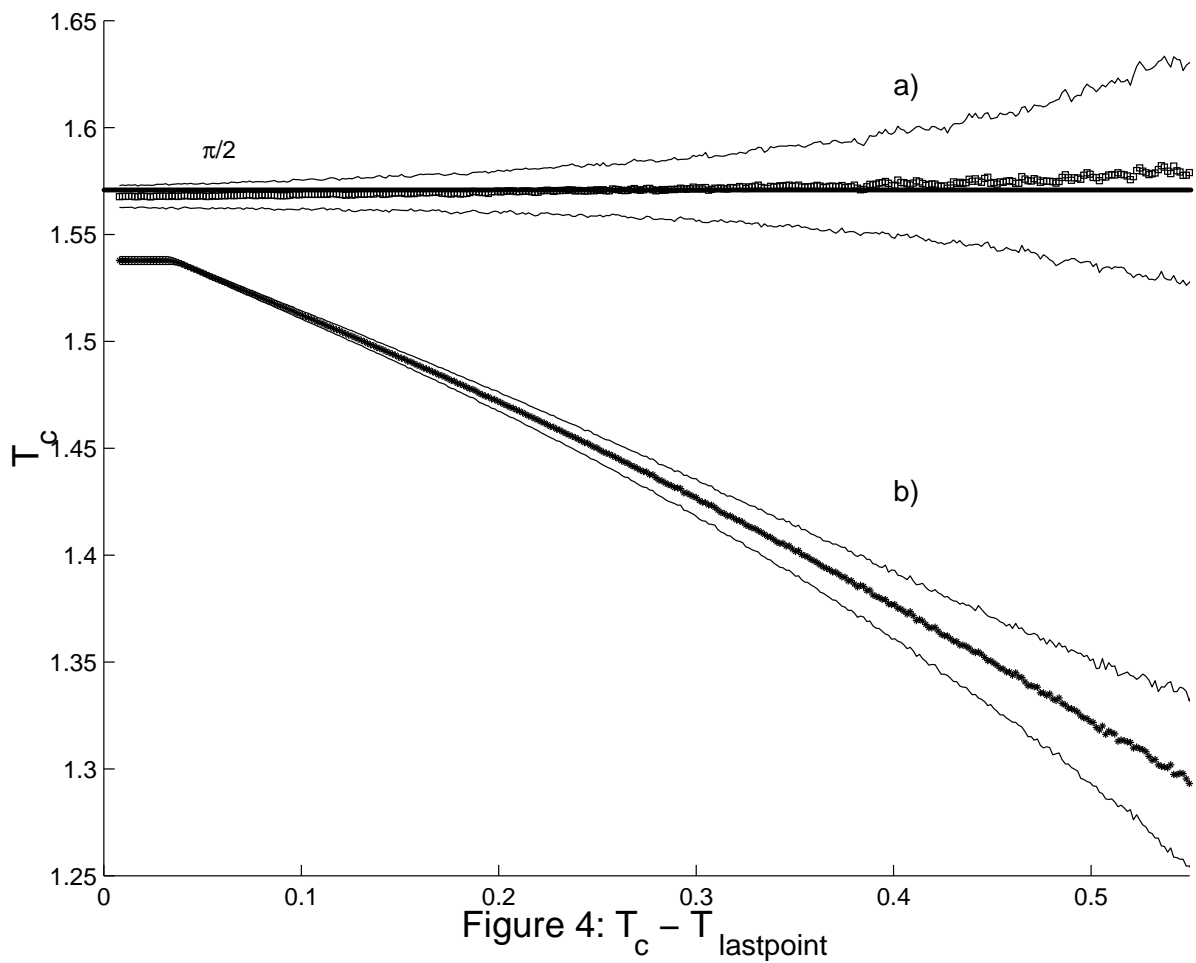,width=16cm}
\caption{\label{fig5} Same as figure 4 
with 1,000 synthetic time series with 
multiplicative noise of variance  $10^{-3}$, and
with different prediction schemes. The curve a) is obtained by using the
general resummation method of section 4.1 which does not assume any specific
value of the exponent $\beta$. We first fit the inverse of each synthetic data
set with a parabolic expression $1+b_1t^2+b_2t^4$ and use these coefficients
$b_1$ and $b_2$ to obtain our prediction $t_c$. The two thin lines are 
the $\pm$ one standard deviations. 
The curve b) is obtained by fitting each data set by the power law equation 
$A (t_c-t)^{-\beta}$, where $A$, $\beta$ and $t_c$ are all three free
parameters determined from the fit
in the time interval $[0.5; T_{\rm lastpoint}]$.
The thick line is the average $t_c$ taken over the 1,000 realizations and the two
thin lines are the $\pm$ one standard deviations.
The horizontal line at $t_c=\pi/2$ is the exact theoretical
value for the critical time. Note the striking superiority of our
resummation method over a direct power law fit.
 }
\end{figure}

Figure \ref{fig5} compares the prediction skill of the Yukalov-Gluzman
method used in section 4.1 to that of a direct power law fit. Again,
1,000 synthetic time series were generated with 
multiplicative noise of variance  $10^{-3}$.
The curve a) of figure  \ref{fig5} is obtained by using the
general resummation method of section 4.1 which does not assume any specific
value of the exponent $\beta$. We first fit the inverse of each synthetic data
set with a parabolic expression $1+b_1t^2+b_2t^4$ and use these coefficients
$b_1$ and $b_2$ to obtain our prediction $t_c$. The two thin lines are 
the $\pm$ one standard deviations. 
The curve b) in figure  \ref{fig5} is obtained by fitting 
each data set by the power law equation 
$A (t_c-t)^{-\beta}$, where $A$, $\beta$ and $t_c$ are all three free
parameters determined from the fit
in the time interval $[0.5; T_{\rm lastpoint}]$.
The thick line is the average $t_c$ taken over the 1,000 realizations and the two
thin lines are the $\pm$ one standard deviations.
The horizontal line at $t_c=\pi/2$ is the exact theoretical
value for the critical time. Note the striking superiority of our
resummation method over a direct power law fit.

\begin{figure}
\epsfig{file=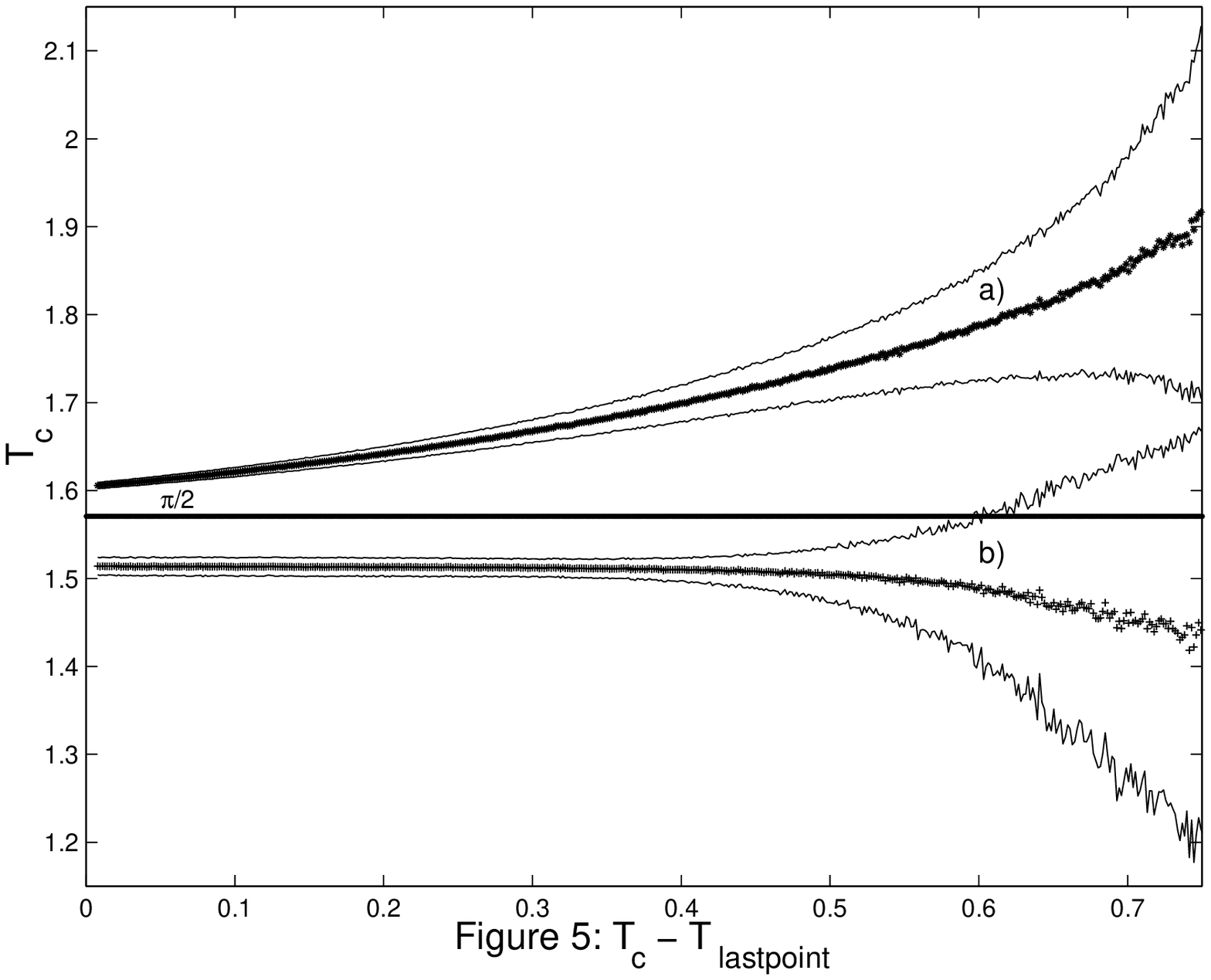,width=16cm}
\caption{\label{fig6} Same as figure 4 but with a
multiplicative noise of variance  $10^{-2}$.
 }
\end{figure}

\begin{figure}
\epsfig{file=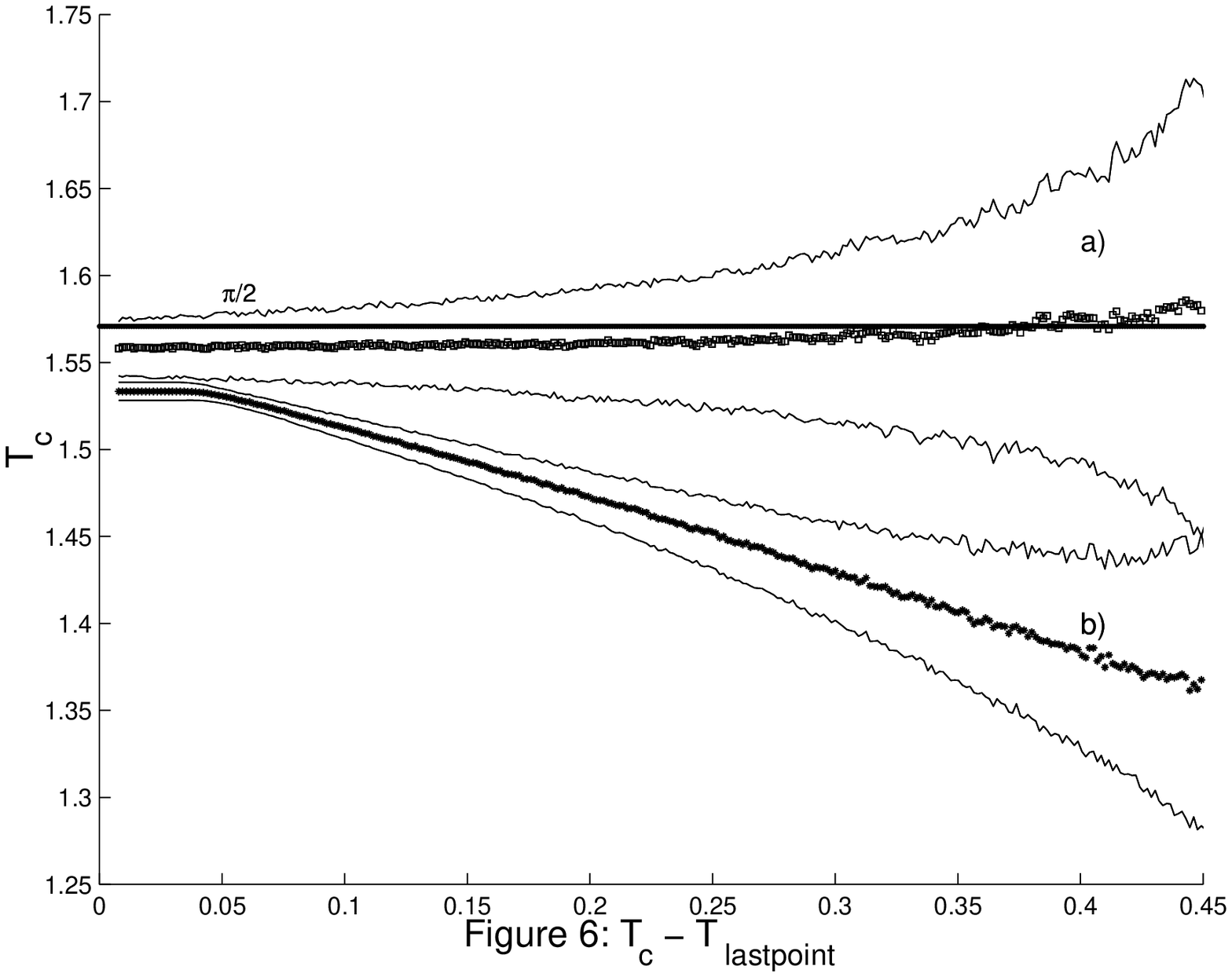,width=16cm}
\caption{\label{fig7} Same as figure 5 but with a
multiplicative noise of variance  $10^{-2}$.
 }
\end{figure}

Figure \ref{fig6} is the same as figure  \ref{fig4} but with a larger
multiplicative noise of variance  $10^{-2}$.
Figure \ref{fig7} is the same as figure  \ref{fig5} but with a larger
multiplicative noise of variance  $10^{-2}$.

\section{Concluding remarks}

We have tested two methods for the prediction of the critical time
of a singular power law behavior and tested their prediction skills
against the direct determination by a power law fit. Our analysis and the
numerical tests convincingly demonstrate the value of our approach which
provides significant improved forecasting skills. Our tests have however
been restricted to a an important from a physical viewpoint but 
still special case of a function which admits
an expansion with only even powers of $t$. The most general situation
contains also odd powers of $t$ which complicates the situation. We intend
to report progress in this general case in a forthcoming publication.

Finally, we wish to comment upon a conceptual understanding of the role of the
exponent $s$ used as control functions in the functional renormalization
schemes used here.
Yukalov and Gluzman \cite{G8} first noticed that their functional
renormalization frameworks allowed them to propose a novel physical
understanding of critical exponents as being directly related to limits of
control functions at the critical point. In other words, they appear as
physical analogs of the rather abstract mathematical objects given by the
control functions. A scale invariant formulation using logarithmic variables
allows us to understand this rather surprising observation: critical
exponents can be seen to be determined by the initial conditions of an
operator of the group of the symmetry of scale invariance \cite{Dubrulle}.
Indeed, a power law function $\phi(r)$ of the distance $r\equiv t_c-t$ to
the critical point, has the property of scale invariance which reads 
\be
\phi(\lambda r)= \lambda^{\alpha} \phi(r)~, \label{vkkaa} 
\ee
where $\lambda$ is an arbitrary magnification factor and $\alpha$ is a
critical exponent. This equation means that the field $\phi(r)$ is invariant
under the homothetical transformation $r \to \lambda r$; $\phi \to
\lambda^{-\alpha} \phi$. Expression (\ref{vkkaa}) can be transformed by
picking up an arbitrary reference field $\phi_0$ and an arbitrary reference
parameter $r_0$ and introducing the log-variables $U = \ln(\phi/\phi_0),
s=r/r_0, U(0)= \ln\left(\phi(r_0)/\phi_0\right), U(s) - U(0)= \alpha s$. In
these variables, equation (\ref{vkkaa}) can also be written: 
\bea
U(s) &=& U(s + \mu) +g(\mu)~, \label{vmnallaa} \\ 
g(\mu) &=& U(0) - U(\mu)~,~~~~~~~{\rm for~any}~~\mu~. \nonumber
\eea
Reciprocally, any regular function obeying (\ref{vmnallaa}) is necessarily
of the shape $U(s) - U(0) = \alpha s$, where 
\be
\alpha = U(1) - U(0) 
\ee
is an arbitrary parameter selected by the initial conditions!

\pagebreak

\end{document}